\batchmode
\makeatletter
\def\input@path{{/Users/YanpengDai/Desktop/Final_Files/}}
\makeatother
\documentclass[10pt,twocolumn,english,conference]{IEEEtran}
\usepackage[T1]{fontenc}
\usepackage{color}
\usepackage{babel}
\usepackage{graphicx}
\usepackage[unicode=true,
 bookmarks=true,bookmarksnumbered=true,bookmarksopen=true,bookmarksopenlevel=1,
 breaklinks=false,pdfborder={0 0 0},pdfborderstyle={},backref=false,colorlinks=false]
 {hyperref}
\hypersetup{pdftitle={Your Title},
 pdfauthor={Your Name},
 pdfpagelayout=OneColumn, pdfnewwindow=true, pdfstartview=XYZ, plainpages=false}

\makeatletter
\IEEEoverridecommandlockouts
\ifCLASSOPTIONcompsoc
\usepackage[caption=false,font=normalsize,labelfont=sf,textfont=sf]{subfig}
\else
\usepackage[caption=false,font=footnotesize]{subfig}
\fi
\usepackage{cite}
\usepackage{amsthm}

\usepackage{algorithm}
\usepackage{algorithmic}
\usepackage{amsmath}
\usepackage{cite}
\usepackage{bm}
\usepackage{algorithmic}
\usepackage{algorithm}
\usepackage{graphicx}
\usepackage{multirow}
\usepackage{subfig}
\usepackage{float}
\usepackage{balance}
\usepackage{setspace}
\usepackage{graphicx}

\@ifundefined{showcaptionsetup}{}{%
 \PassOptionsToPackage{caption=false}{subfig}}
\usepackage{subfig}
\makeatother

\begin{document}

\title{Big Data Driven Vehicular Networks}

\author{\IEEEauthorblockN{Nan~Cheng, Feng~Lyu, Jiayin~Chen, Wenchao
Xu, Haibo Zhou, Shan Zhang, Xuemin (Sherman) Shen}\thanks{Nan Cheng, Jiayin Chen, Wenchao Xu, and Xuemin (Sherman) Shen are with the Electrical and Computer Engineering Department, University of Waterloo, Waterloo, ON, N2L 3G1, Canada (emails:\{n5cheng,j648chen,w74xu,sshen\}@uwaterloo.ca)}
\thanks{Feng Lyu is with the Department of Computer Science and Engineering, Shanghai Jiao Tong University, Shanghai, 200240, P.R. China. (email: fenglv@sjtu.edu.cn)}
\thanks{Haibo Zhou is with the School of Electronic Science and Engineering, Nanjing University, Nanjing, 210023, P.R. China  (email: haibozhouuw@gmail.com). Haibo Zhou is the corresponding author.}
\thanks{Shan Zhang is with the Department of Computer Science and Technology, Beihang University, Beijing, 100083, P.R. China (email: zhangshan\_2011@outlook.com)}}
\maketitle
\begin{abstract}
Vehicular communications networks (VANETs) enable information exchange
among vehicles, other end devices and public networks, which plays
a key role in road safety/infotainment, intelligent transportation
system, and self-driving system. As the vehicular connectivity soars,
and new on-road mobile applications and technologies emerge, VANETs
are generating an ever-increasing amount of data, requiring fast and
reliable transmissions through VANETs. On the other hand, a variety
of VANETs related data can be analyzed and utilized to improve the
performance of VANETs. In this article, we first review the VANETs
technologies to efficiently and reliably transmit the big data. Then,
the methods employing big data for studying VANETs characteristics
and improving VANETs performance are discussed. Furthermore, we present
a case study where machine learning schemes are applied to analyze
the VANETs measurement data for efficiently detecting negative communication
conditions.
\end{abstract}

\section{Introduction\label{sec:Intro}}

With the development of automobile technologies, vehicles are expected
to be not only safer, but also greener, more comfortable and entertaining,
while self-driving is also a defining requirement of the future vehicles.
As a promising technology to meet such expectations, vehicular communication
networks (VANETs) \textcolor{black}{enable automobiles to communicate
with each other through vehicle-to-vehicle (V2V) communication and
the network through vehicle-to-infrastructure (V2I) communication,
and exchange information efficiently and reliably through the V2V
and V2I communications, or more generally, vehicle-to-everything (V2X)
communications.} VANETs can facilitate a variety of useful applications,
such as road safety enhancement, traffic management, vehicular mobile
data services, and self-driving assistance \cite{al2014comprehensive,lu2014connected}.

Due to the ever-increasing demand of mobile services and the fast
development of self-driving technologies, the data volume required,
generated, collected, and transmitted by VANETs has seen an exponential
escalation, which is known as big data \cite{chen2014big}. As explained
in \cite{bedi2014use}, the data in VANETs can well match the ``5Vs''
of big data characteristics, i.e., volume, variety, velocity, value,
and veracity, which justifies that the VANETs data can be treated
as big data and can be solved by big data techniques.

\begin{figure*}
\centering\includegraphics[height=7cm]{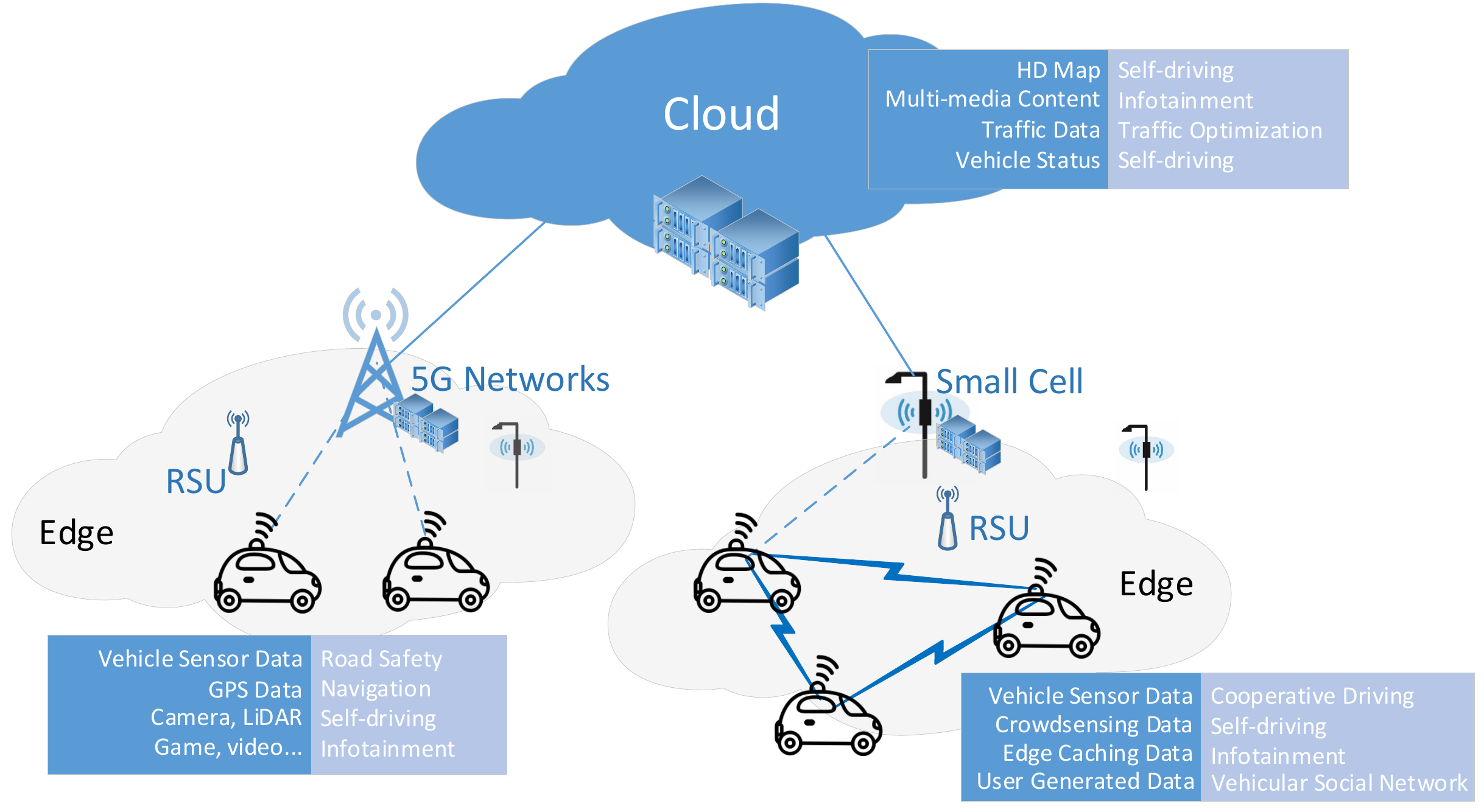}

\caption{VANETs big data and applications.}

\label{Fig:big_data_VANET}
\end{figure*}

Relying on the big data, th\textcolor{black}{e future VANETs will
enable a variety of promising applications and services, such as smart
city and Intelligent Transportation System (ITS) applications, and
significantly change many aspects of the society, including the transportation
system, telecommunication, business, government, as well as the human
life style. The VANETs big data and enabled applications are shown in Fig.
\ref{Fig:big_data_VANET}. For example, road traffic information can
be collected by vehicles and road-side units, and reported to the
ITS cloud server. Based on large-scale traffic information, real-time
traffic prediction and management functions are conducted, so as to
detect the road anomaly, alleviate traffic jam, and reduce emission
and pollution. Self-driving vehicles will consume or generate multiple
Giga Bytes (GB) data per second, typically from outfitted high-quality
cameras, LiDARs and Radars \cite{maurer2016autonomous}. Through the
data fusion, analysis and integration of the cloud data such as weather
and road traffic information, and information from other vehicles,
self-driving vehicles can make decisions on actuating the vehicle
for driving autonomously, on a planned route, and eliminate traffic
fatalities. As a potential impact of self-driving technologies, the
vehicles will be more like home or offices, and thus people will focus
on the mobile applications and services that can better support the
in-vehicle activities, rather than driving the vehicle. Therefore,
the future VANETs will evolve to satisfy the big mobile data demands,
and support a wide variety of promising applications and services.}

\textcolor{black}{The trend of big data can bring new challenges and
opportunities for VANETs. On one hand, the VANETs big data is with
a significantly large amount, from heterogeneous sources, and having
various requirements. To efficiently support the big data, VANETs
should be capable of providing extremely high data rate, large network
capacity, heterogeneous network integration, and differentiated quality
of service (QoS) guarantee. In addition, besides data communication,
the future VANETs are envisioned to play a critical role in data collection,
storage, and computation. On the other hand, the VANETs big data such
as GPS, vehicle mobility trace, road traffic information, and network
measurements, contains rich valuable network information. If properly
utilized, such big data can reveal a lot of network characterizations,
evaluate the network performance, and optimize the network management,
by applying advanced techniques such as big data mining, analysis,
and machine learning mechanisms. The purpose of this article is to
investigate the impacts of big data on VANETs, introduce the new challenges
and opportunities, and discuss corresponding solutions. We focus on
two related topics, i.e., efficiently supporting big data in VANETs,
and utilizing big data for better understanding and improving VANETs.
Furthermore, we study a case where machine }learning schemes are applied
to analyze the VANETs measurement data for efficiently detecting negative
communication conditions.

\section{Big Data in VANETs\label{sec:BigData}}

The VANETs big data come from multiple heterogeneous sources, presenting
diversified characteristics, such as volume, structure, value, requirements
for processing delay, etc. We classify the VANETs big data according
to the sources of the data as follows.
\begin{itemize}
\item \textbf{Vehicle sensing data}: Modern vehicles have equipped various
sensors (speedometer, tire pressure sensor, etc.) to collect vehicle
and environmental information. Rich information from such sensors
can enable a wide range of applications, such as online vehicle diagnosis,
road safety improvement, smart charging planning, accident detection,
and so forth.
\item \textbf{GPS data:} GPS devices can provide accurate and structured
location-related information of vehicles, including longitude, latitude,
altitude, and speed. GPS data can be used for diversified goals, such
as navigation, traffic management, communication routing optimization,
vehicular content caching and sharing, etc. In addition, the datasets
of large-scale vehicle trajectories, generated by tracing the long-time
GPS data of vehicles in a geographical area, can be investigated to
analyze the VANETs characteristics, such as network connectivity,
and design efficient mechanisms, such as routing protocol for delay-tolerant
vehicular network, and radio access network deployment.
\item \textbf{Self-driving related data:} The autonomous vehicle will make
big data even bigger. Self-driving technology requires the accurate
perception and understanding of the environment to make proper decisions
to control the vehicle. Since traditional sensors have limited capability,
and cannot provide necessary information such as real-time road vision,
accurate distance, and 3D map, advanced devices like cameras and light
detection and ranging (LiDAR) sensors are equipped for a better perception.
However, the high-definition cameras and LiDAR will produce a huge
amount of data as they continuously collect high-definition data such
as high-quality videos.
\item \textbf{Vehicular mobile service data:} In-vehicle infotainment is
becoming more crucial for improving the experience of both drivers
and passengers. Mobile applications such as video/audio streaming,
online gaming, social networks, and user generated contents (UGC)
require or generate a huge amount of data.
\end{itemize}

\section{Supporting Big Data in Vehicular Networks\label{sec:Supporting}}

For big data system to efficiently function, four essential parts
need to be well supported, i.e., data aggregation, storage, tr\textcolor{black}{ansmission,
and computation. In VANETs, the raw data can be gathered by vehicle
sensors, and stored in on-board storage. Since the raw data contain
redundancy, data processing is conducted to extract valuable information.
After accumulating the data (either raw or processed), there is a
demand to transmit the data to appropriate data storage systems (such
as cloud/edge servers) for further analysis and process. Therefore,
VANETs should be capable of effectively supporting these big data
functions. }

\textcolor{black}{}
\begin{figure*}
\textcolor{black}{\centering\includegraphics[height=8cm]{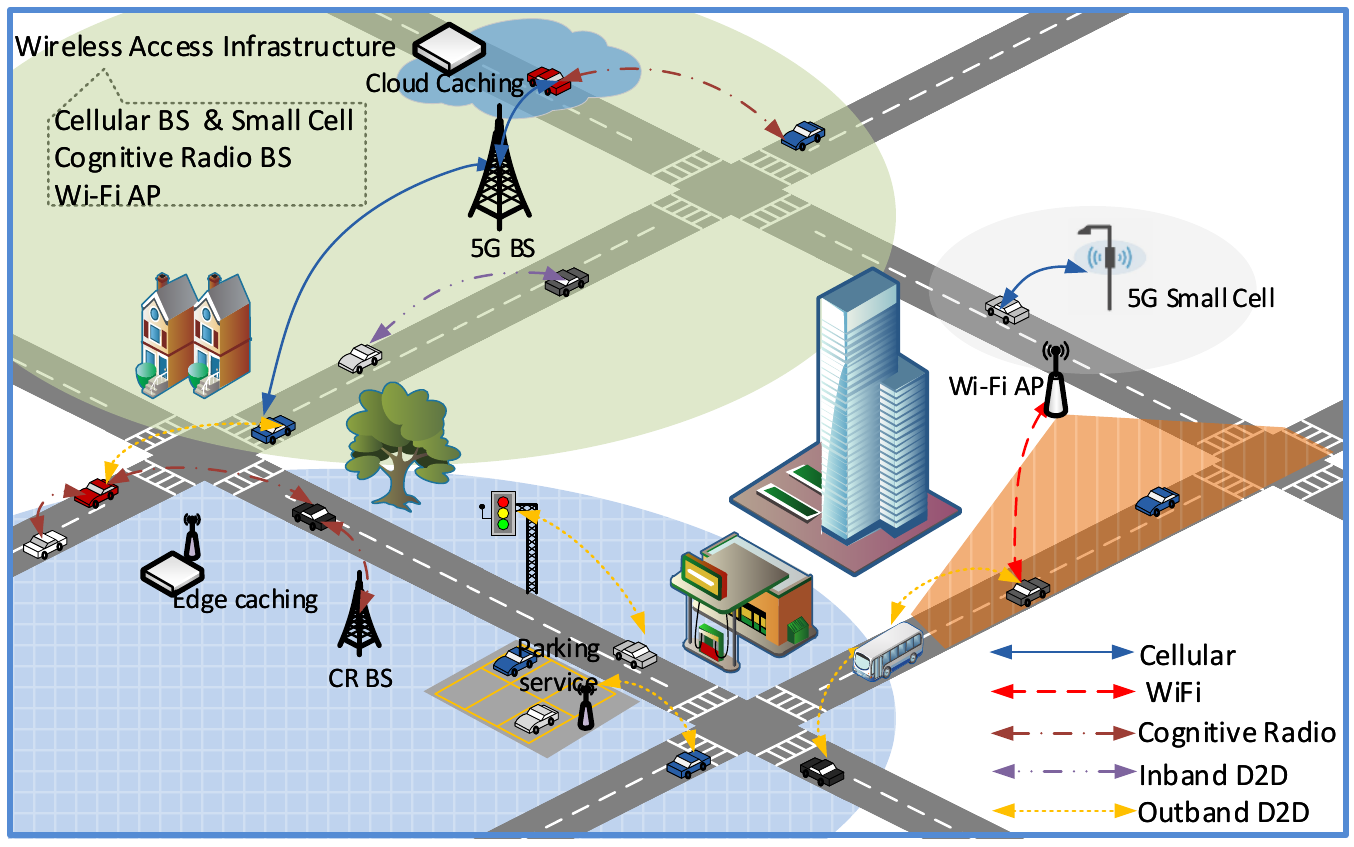}}

\textcolor{black}{\caption{Supporting Big Data Through VANETs}
}

\textcolor{black}{\label{fig:support_big_data}}
\end{figure*}

\textcolor{black}{Traditional VANETs employ the IEEE 802.11p based
dedicated short-range communication (DSRC) technologies, where data
transmission mainly relies on distributed medium access control (MAC)
and multi-hop routing protocols \cite{hadded2015tdma}. However, the
traditional VANETs technologies can hardly satisfy the harsh requirements
of big data applications due to the decentralized protocols and bandwidth
limitations, which leads to the lack of network resources and flexibility
to support the big data with diversified QoS requirements. Moreover,
issues such as energy efficiency, caching, and computation capabilities
are not well considered in current VANETs, which are also essential
in supporting the big data. In this section, we discuss some promising
VANETs technologies to better support the big data, including 5G technologies
and opportunistic data offloading mechanisms. As shown in Fig. \ref{fig:support_big_data},
the 5G macro cells can provide ubiquitous communication support, while
5G small cells, Wireless Local Area Networks (WLANs), cognitive radio
networks (CRNs) and device-to-device (D2D) communications offer cost-effective
data pipes for VANETs big data.}

\subsection{5G Technologies}

An intuitive solution to support the VANETs big data is the pervasive
cellular network. As the 4G LTE network is struggling to support the
ever-increasing data volume and the emerging mobile services with
differentiated QoS requirements, 5G networks, the next-generation
networks, are building a way to solve the issues. Based on software-defined
network (SDN) related technologies, 5G networks are designed to serve
as a platform to provide satisfying services for vertical fields,
including telecommunication, transportation, agriculture, economics,
government, education, etc \cite{quan2017enhancing}. According to
the key performance indicators, 5G networks are capable of offering
a 10 Gb/s data rate with less than 1 ms end-to-end latency \cite{andrews2014will}.
Moreover, machine-type communications with low power consumption and
high reliability requirements are well supported for the emerging
Internet of Things (IoT) applications.

To better characterize and support different services, 5G defines
three categories of use cases, i.e., enhanced mobile broadband (eMBB),
ultra-reliable and low-latency communication (URLLC), and massive
machine-type communication (mMTC), and the performance indicators
of each categories. These three categories, together with the well-defined
key technologies, can provide guaranteed performance to VANETs big
data gathering and transmission tasks.
\begin{itemize}
\item eMBB: In VANETs, the exponentially increasing big data demands of
the vehicular mobile data services requires a high-capacity network
that can provide extremely high date rates. Enabled by promising network
technologies, such as advanced channel coding, mmWAVE, and ultra-dense
small cell networks, eMBB can provide peak data rate of 10 Gb/s and
mobile data volume of 10 Tb/s/km$^2$. Therefore, with 5G networks,
the emerging data-craving vehicular data applications can be better
supported, and many more will come to reality.
\item \textcolor{black}{URLLC: The mission-critical data services in VANETs,
such as safety message transmission, require very low latency and
very high reliability. The requirements fall into the category of
URLLC in 5G, which can provide less than 5 ms latency and higher than
99.999\% reliability. }
\item \textcolor{black}{mMTC: Relying on potential technologies such as
machine-to-machine communication and narrow-band IoT (NB-IoT), mMTC
aims to support ubiquitous machine-type connections with low energy
consumption and low latency. A large amount of VANETs big data is
generated by the densely deployed light weight devices, such as sensors
equipped in vehicles or deployed along the roads. 5G technologies
can accommodate such massive concurrent connectivity, provide reliable
data transmission, and prolong the device battery life, and therefore
facilitate the big data gathering services.}
\end{itemize}
\textcolor{black}{5G also defines }enhanced vehicle-to-everything
(eV2X) use case for supporting the vertical field of vehicular communication
and data services \cite{3gppenhance}. The requirements for typical
V2X scenarios are defined, including vehicle platooning, advanced
driving, extended sensors, and remote driving.

\subsection{Opportunistic Data Pipes}

Although the 5G networks can significantly improve the network capacity,
the ever-increasing big data will still put a severe burden on the
network, resulting in possible network congestions. In addition, the
commercialization and deployment of 5G networks will start in year
2020, and will be a long-time process. Therefore, in the near future,
the 4G LTE networks with relatively small capacity will be straining
to accommodate the big data. Moreover, usually using the cellular
network to transmit a large amount of data will incur prohibitive
costs. As a result, alternative data pipes for supporting the big
data are required. WLANs, CRNs and D2D communications can be employed
to offload the VANETs big data from the cellular network in a cost-effective
way.

\subsubsection{WiFi Offloading}

WiFi, operating on unlicensed spectrum, is a popular solution to deliver
data content at low cost. The feasibility of WiFi for outdoor Internet
access at vehicular mobility, referred to as drive-thru Internet,
has been demonstrated in \cite{bychkovsky2006measurementfull}. Different
from the fully covered cellular network, WiFi only provides intermittent
small coverage areas along the road. Therefore, although WiFi operates
on unlicensed spectrum, it is spatially/temporal opportunistic for
vehicles to employ due to the vehicle mobility. Therefore, employing
the mobility feature is an important issue in vehicular WiFi offloading.
One example is prediction-based delayed offloading. Based on the mobility
prediction and priori knowledge of WiFi deployment, the future opportunities
of WiFi access and corresponding throughput can be predicted. Then,
according to the delay tolerance of different users or applications,
offloading decision can be made whether to wait for WiFi offloading
or directly transmit through cellular networks.

\subsubsection{Cognitive Radio Technology}

Cognitive radio is envisioned as a promising spectrum-sharing technology
which enables unlicensed users opportunistically exploit spatially
and/or temporally vacant licensed radio spectrum bands which are allocated
to licensed systems. The CR technology can employ the vast underutilized
spectrum resources to support the big data transmissions. However,
in VANETs, the high mobility of vehicles may require excessively frequent
spectrum sensing to protect the primary transmissions \cite{cheng2013vehicular}.
The TV white spaces (TVWS) have been suggested for wireless broadband
access due to the abundant and currently underutilized spectrum resources
at VHF/UHF bands and its superb penetration property. Unlike other
licensed system, the spectrum usage of TV broadcasting system is highly
stable and predictable, and can be inquired from a database. Therefore,
the TVWS is envisioned as a potential solution to CR-enabled VANETs
\cite{zhou2017tv}.

\subsubsection{Device-to-Device Communication}

By utilizing the proximity, mobile users can communicate directly
with each other using the cellular spectrum (or other spectrum bands)
without traversing the base station or the backhaul networks, named
device-to-device D2D communications. Therefore, D2D communications
can increase the overall spectral efficiency and reduce communication
delay for mobile users, which may be applied to many VANETs applications
such as video streaming, location-aware advertisement, safety related
applications, and so forth. However, incorporating D2D communication
in vehicular environment introduces several new challenges. For example,
a full channel state information, which is usually needed in resource
allocation schemes for D2D communication, is hard to track and easy
to be outdated in VANETs. In addition, the topology of VANETs makes
the interference pattern more difficult to model than a general cellular
network where a Poison point process (P.P.P.) can be applied to model
the user spatial distribution.

\section{Employing Big Data in Vehicular Networks\label{sec:Employing}}

\begin{figure*}
\centering\includegraphics[height=8cm]{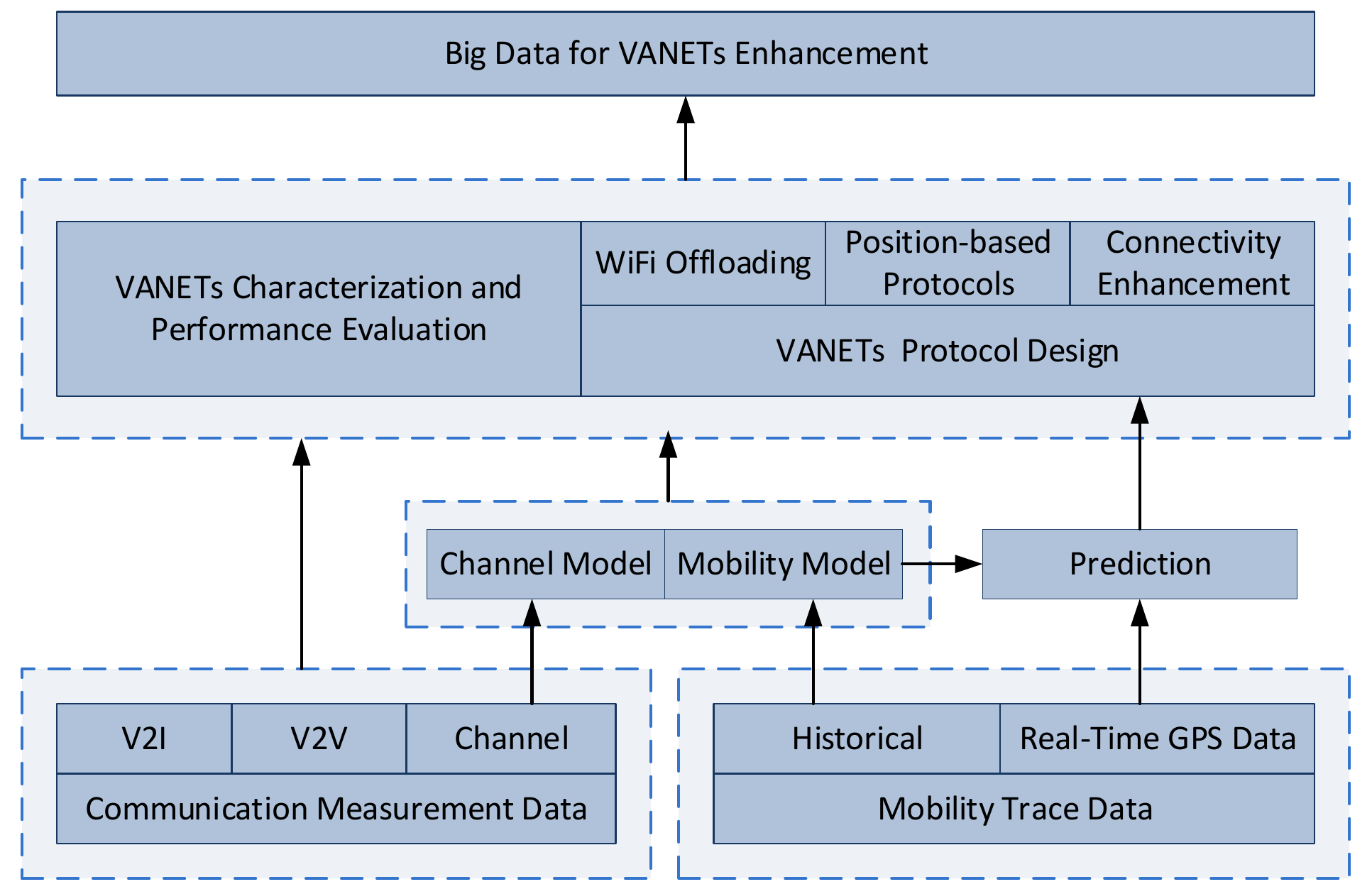}

\caption{Employing big data in VANETs.}

\label{fig:employing}
\end{figure*}

As mentioned above, big data \textcolor{black}{in VANETs can provide
valuable insights of VANETs, which can be employed to characterize
and evaluate the performance of VANETs, and design new protocols with
big data intelligence. In this section, we show the utilization of
two typical data sets in VANETs, i.e., vehicle mobility trace data
and VANETs measurements data. An overview of big data employment in
VANETs is shown in Fig. \ref{fig:employing}. The two data sets can
be employed to extract practical channel model and mobility model,
and predict vehicle movement. With such knowledge, VANETs characterization
and intelligent protocol design can be achieved. }

\subsection{\textcolor{black}{Vehicle Mobility Trace Data}}

\textcolor{black}{Admittedly, the high mobility of vehicles leads
to challenges to VANETs. However, the mobility can also have benefits
on the network, e.g., mobility-aware protocols and delay-tolerant
data dissemination. Through the analysis of the datasets} of vehicle
mobility, an amount of valuable information can be obtained, such
as the practical mobility model, network connectivity, spatial and
temporal density distribution, etc. There are several databases that
stores real and large-scale taxi mobility trace data from different
cities, including San Francisco, Shanghai, and Shenzhen \cite{celes2017improving}.
Main content of the trace data includes time stamp, vehicle velocity,
driving direction and vehicle location, which can be used for further
study on VANETs.

Mobility model is widely used in VANETs location-based protocol design
and performance evaluation. Due to the time intervals of vehicles
reporting their trace, the trace data is always error-prone and has
gaps between locations in two consecutive records. Therefore, some
data preprocessing mechanism is needed. For instance, due to the predictability
property of vehicle mobility, it is possible to fill the gap by predicting
the route through analyzing road map, traffic signs and the past vehicle
trace. Then, a realistic mobility model can be generated from the
modified trace data.

Position-based routing schemes and MAC protocols are designed to adapt
to the high mobility and frequently changing topology of VANETs. The
mobility model and network characteristics can be obtained by analyzing
the mobility trace data and network measurement data, which are taken
into consideration in the design of routing schemes and MAC protocols.
For instance, position-based routing schemes can exploit the real-time
position and predict vehicle movement to improve the transmission
performance. Position-based MAC protocols can predict potential packet
collisions due to the vehicle mobility and make effort to avoid them.
The historical mobility trace data can also be used in simulations
to evaluate the designed MAC and routing protocols.

Furthermore, mobility trace data is also useful in analyzing and improving
the connectivity of VANETs. Network connectivity metrics can be evaluated
from the mobility trace data, including link duration, average hops,
number of connected vehicle pairs, and interconnect time distribution.
Improvement of connectivity can also be achieved with the aid of trace
data. The prediction methods of vehicle movement can be developed
to make seamless handoff possible for communication between vehicles
and infrastructures. In addition, through investigating the real-time
trace data generated in VANETs, information of vehicle traffic flow
can be obtained. Then, unmanned aerial vehicles (UAVs) can be deployed
in order to improve the network connectivity.

\subsection{VANETs Measurement Data}

Measurement of VANETs communication plays a vital role in VANETs characterization,
since in VANETs, many influencing factors are difficult to model,
such as mobile channels, pedestrians, terrain, and obstacles. In order
to obtain realistic measurement data, communication devices using
IEEE 802.11p protocol are deployed on vehicles and road-side units
(RSUs) during experiment. These experiments are conducted in various
environments such as urban, suburban, rural, open fields and freeway,
and different measurement data is collected depending on the characteristics
of interest.

WiFi offloading is envisioned as a potential solution to data explosion
problem in cellular networks. However, high mobility of vehicles makes
WiFi offloading in VANETs distinguished from static WiFi offloading.
Measurement data like connection establishment time, connection time,
interconnection time, max rate and transferable data volume in once
drive-thru is collected to analyze WiFi offloading performance. Then,
a three-phase feature is observed as an important WiFi offloading
characteristic, including entry, production, and exit phases. It shows
that in entry and exit phases, the connection quality is weaker and
data rate is lower than production phase, which provides guidance
to researchers about how to improve the offloading performance, e.g.,
reducing the association and authentication time in order to maximize
data transfer in production phase.

Unlike static or low-mobility wireless channels, the vehicular channel
is more complicated due to the shadowing by nearby vehicles, high
Doppler shifts, and inherent nonstationary \cite{mecklenbrauker2011vehicular}.
Therefore, building an accurate and practical channel model is crucial
for VANETs performance analysis, protocol design, and simulation experiments.
This can be done by studying the real communication measurement data,
including both V2V measurements and V2I measurements in different
important environments. The resulting channel models characterize
the vehicular channel from different channel metrics, including pathloss,
signal fading, delay spread, Doppler spread, and angular spread.

\section{Case Study\label{sec:Case}}

In this section, we study a case where big data and machine learning
schemes are employed to support efficient protocol design in VANETs
communications.

\subsection{Online NLoS Detection}

In VANETs, packets related to safety information should be delivered
perfectly (with transmission chance and without packet loss). However,
it is found that non-line-of-sight (NLoS) condition is a key factor
of V2V link performance degradation \cite{lv2016empirical}. Inspired
by the intuition that blindly sending more packets in harsh NLoS conditions
can hardly succeed but incur resource wasting and increase interference
to other neighboring vehicles, we propose an innovative scheme to
\emph{detect NLoS conditions online} by learning the V2V measurement
data. Given that the NLoS condition can be detected, more robust protocols
can be devised, e.g., allocating scarce wireless channel resources
to those vehicles under line-of-sight (LoS) conditions or seeking
helper vehicles to relay packets for those vehicles under NLoS conditions.
In the sequel, we will elaborate the scheme in two parts, i.e., measurement
data collection and building detection model using machine learning
methods.

\begin{figure*}[t]
\subfloat[Data collection devices]{\includegraphics[height=3.5cm]{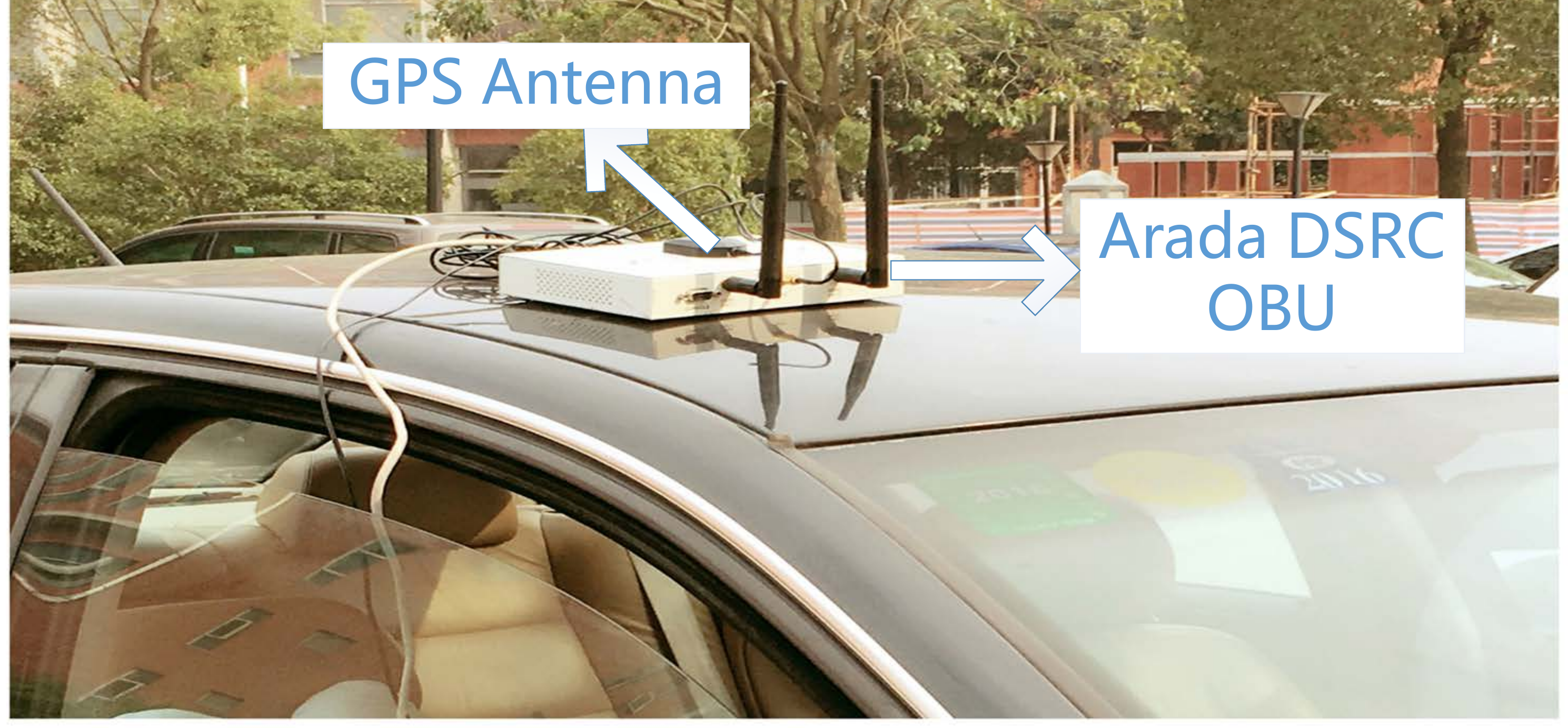}}\hspace{1.8cm}\subfloat[Data collection scenarios]{\includegraphics[height=3.5cm]{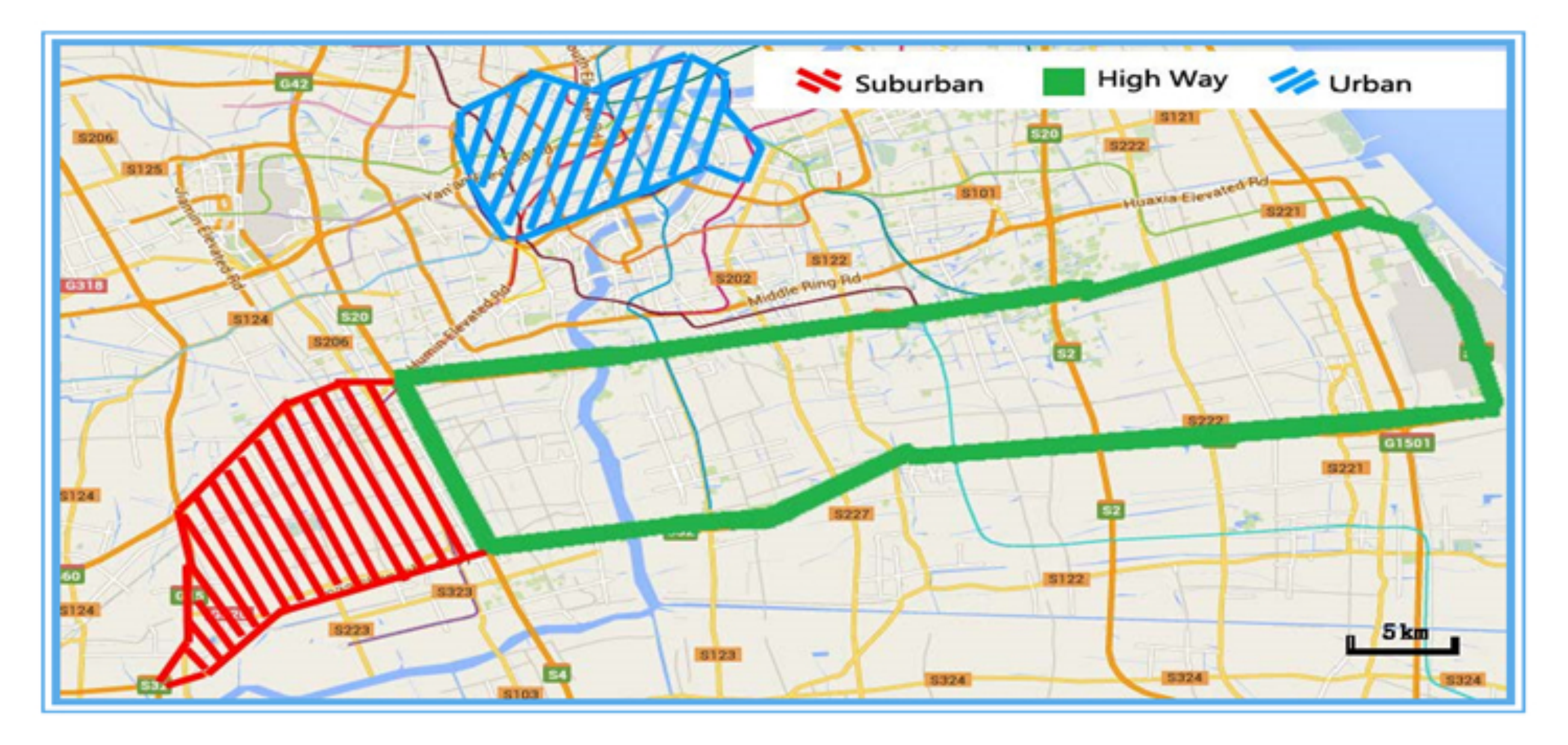}}

\subfloat[A LoS condition]{\includegraphics[height=3.5cm]{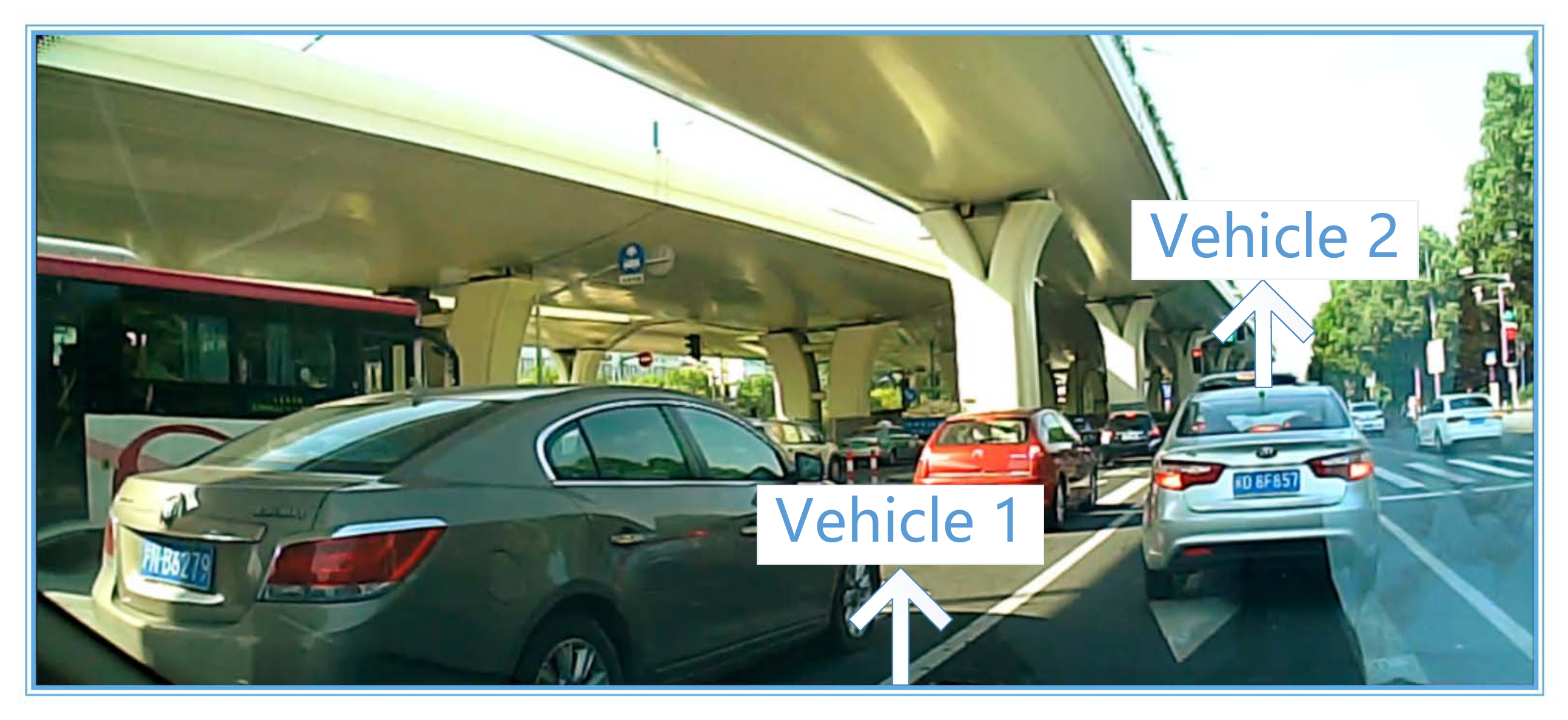}}\hspace{1.5cm}\subfloat[A NLoS condition]{\includegraphics[height=3.5cm]{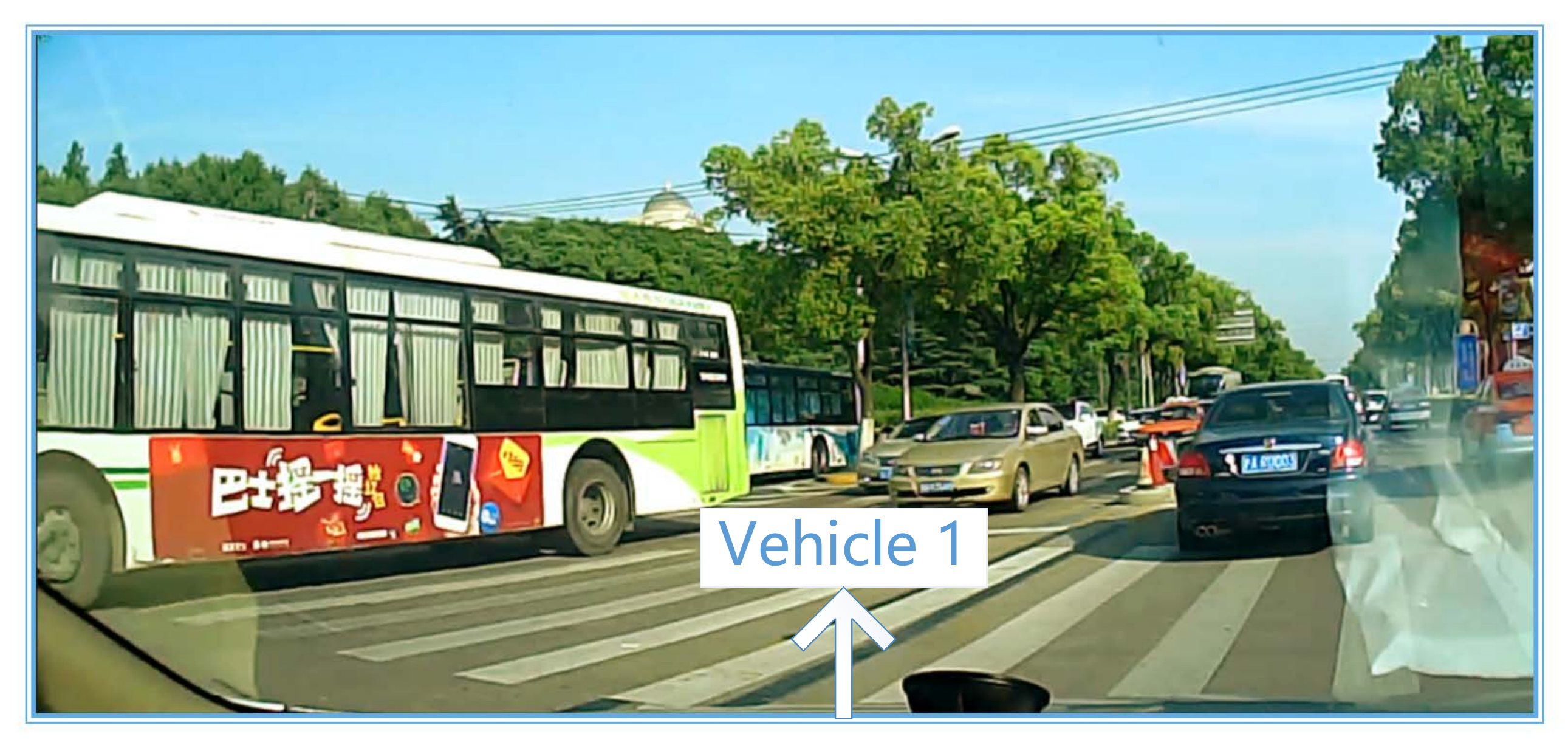}}

\caption{\textcolor{black}{Illustration of the data collection campaign}}
\label{fig:DSRC_illu}
\end{figure*}

\subsection{Collecting V2V Communication Measurement Data Sets}

We collect V2V communication trace data by two experimental vehicles
each mounted with a Arada LocoMateTM OBU (DSRC module) on the roof.
The transmitter vehicle sends a 300-bytes packet every 100 ms to the
receiver vehicle, which consists of a sequence number, the latitude,
the longitude, the altitude and the speed information of the transmitter.
Meanwhile, both the transmitter and the receiver log all the packets
transmitted and received. In addition, we deploy two cameras on each
vehicle with one mounted on the front glass and the other fixed on
the rear glass, which record the whole process for off-line analysis.
Fig. \ref{fig:DSRC_illu} (a) shows the data collection devices.

We conduct data collection campaigns including three major road types
in a city, i.e., highway, suburban, and urban. Each data set contains
following three types information: 1) communication trace: by comparing
packet's sequence number at sender and receiver, each packet can be
marked as received or dropped and we can compute the packet delivery
ratio (PDR) throughout all experiment time; 2) GPS trace: both vehicles
have logged GPS trace, which can provide speed, altitude and distance
information; 3) recorded videos: it can be utilized to check the communicating
environments, e.g., types of road, traffic conditions, surrounding
obstacles and so on. Three types of data are within time synchronization
for better observation and comparison. The overall campaign lasts
for over two months with an accumulated distance of over 1,500 kilometers
and a total size up to 110GB. We run our testbed within areas of the
above three road types in Shanghai as shown in Fig. \ref{fig:DSRC_illu}
(b). We denote three data sets by $\mathcal{H}$ (highway), $\mathcal{S}$
(suburban), and $\mathcal{U}$ (urban).

\subsection{Supervised Machine Learning}

In this subsection, we use two classic supervised machine learning
methods, i.e., \emph{Naive Bayes (NB)} and \emph{Support Vector Machines (SVM)},
to detect NLoS conditions.

\textbf{Labeling NLoS conditions:} Before using machine learning
techniques, we first label out all NLoS conditions. Since the whole
data collection campaigns are recorded by cameras, we mark all NLoS
situations when two vehicles cannot visually see each other. Although
NLoS conditions found by cameras are not necessarily to be NLoS for
RF radios, those visually NLoS conditions are still good approximations
of real radio NLoS conditions and valuable for learning. Fig. \ref{fig:DSRC_illu}(c)
and Fig. \ref{fig:DSRC_illu}(d) show examples of a LoS condition
and a NLoS condition, where vehicle 1 and vehicle 2 are communicating
vehicles, but vehicle 2 is blocked by obstacles in the NLoS condition
and cannot be found in Fig. \ref{fig:DSRC_illu}(d).

\begin{figure*}[t]
\subfloat[Accuracy under different scenarios]{\includegraphics[height=6.5cm]{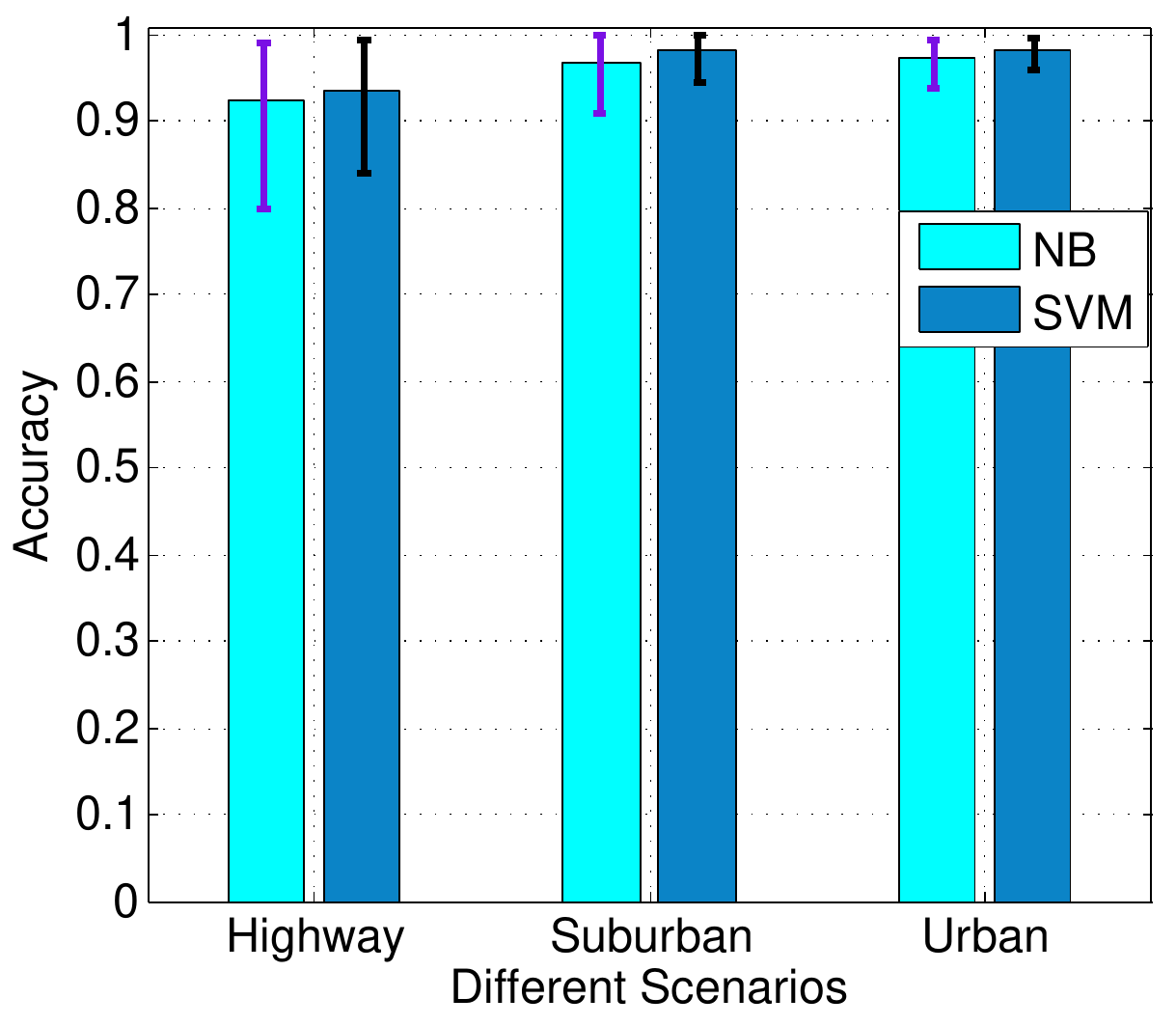}}\hspace{1.8cm}\subfloat[Accuracy v.s. traing data size.]{\includegraphics[height=6.5cm]{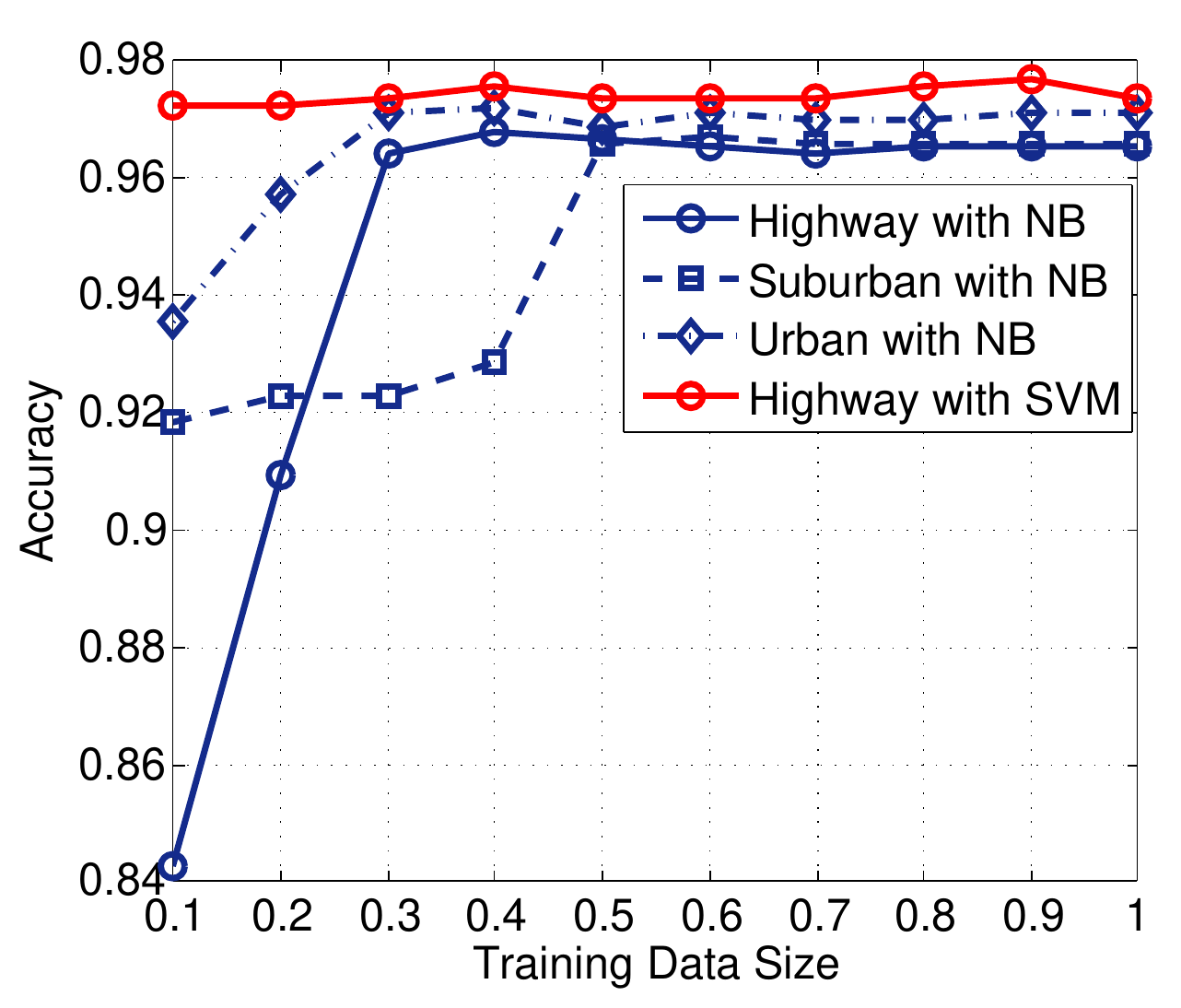}}

\caption{NLoS detection accuracy.}

\label{fig:accu}
\end{figure*}

\textbf{Feature Extraction:} When machine learning algorithms are
processed, representative tuple of features rather than raw data is
a more effective input. Thus, it is necessary to extract effective
features from raw data set. According to the analysis in the work
\cite{lv2016empirical}, PDRs are heavily influenced by LoS/NLoS conditions
and LoS/NLoS durations are with memories due to the power law distributions.
Therefore, we can use history PDR values as features for training.
At this point, we select three features, i.e., PDR value of the previous
1 second, PDR value of the previous 5 seconds and PDR value of the
previous 10 seconds.

\textbf{Machine learning with NB and SVM:} After feature extracting,
we obtain samples in the form of <3-dimensional features, label>.
We then use parts of samples to train NB and SVM models. NB methods
are a set of supervised learning algorithms based on applying Bayes\textquoteright{}
theorem. Given a label variable $y$ and a tuple of feature vectors
$x_1$ to $x_n$, Maximum A Posteriori (MAP) estimation is used to
estimate $P(y)$ and $P(x_i|y)$. NB learners and classifiers can
be extremely fast compared to some sophisticated methods. The cores
in SVM are the kernel and the similarity function. A kernel is a landmark,
and the similarity function computes the similarity between an input
example and the kernels.

\subsection{Performance Evaluation}

To evaluate the performance of machine learning methods, we define
the following metrics based on the True Positive (TP), True Negative
(TN), False Positive (FP) and False Negative (FN): 1) \emph{accuracy:}
the probability that the identification of a condition is the same
as the ground truth; 2) \emph{precision:} the probability that the
identifications for NLoS conditions are exactly NLoS conditions in
ground truth; 3) \emph{recall:} the probability that all NLoS conditions
in ground truth are identified as NLoS conditions; 4) \emph{false positive rate (FPR):}
the probability that a LoS condition is identified as a NLoS condition.

We first evaluate the learning results under different scenarios.
For more robust model evaluation, we adopt the cross-validation scheme
to validate training models. In specific, for each data set, i.e.,
$\mathcal{H}$ with 16425 samples, $\mathcal{S}$ with 16033 samples
and $\mathcal{U}$ with 27439 samples, we first split them to 10 subsets,
then cross validate the learning models by using $i$-th subset, for
$i \in \{1, 2, ..., 10\}$, as validation set and the remaining subsets
together as training sets. Fig. \ref{fig:accu}(a) shows the accuracy
of NLoS detection under different scenarios with NB and SVM methods.
We have the following two main observations. First, both NB and SVM
methods can achieve superb accuracy values. For instance, with NB
method, the accuracy can reach about 92.5\%, 96.9\% and 97.4\% in
highway, suburban and urban, respectively, while with SVM, the values
can be about 93.7\%, 98.3\% and 98.3\%, respectively. Second, the
performance of SVM can slightly outperform the performance of NB.
Table I shows other metric values and similar observations can be
obtained.

\begin{table*}[!t]  \renewcommand\thetable{\Roman{table}} \small \setlength{\arrayrulewidth}{0.2mm} \begin{center} \caption{Learning results} \begin{tabular}{c c c c c c c c c}   \hline   \multirow{2}{*}{Scenarios}   &\multicolumn{2}{c}{ \underline{Accuracy(\%)}}  & \multicolumn{2}{c}{\underline{Precision(\%)}} & \multicolumn{2}{c}{\underline{Recall(\%)}}  & \multicolumn{2}{c}{\underline{FPR(\%)}} \\   &NB & SVM &NB & SVM &NB & SVM &NB & SVM\\   \hline   Highway & 0.9247 & 0.9367 & 0.9578& 0.9393 & 0.9359 & 0.9832 & 0.0606 & 0.0189 \\   Suburban & 0.9690 & 0.9831 & 0.9958& 0.9867 & 0.9715 & 0.9943 & 0.0280 & 0.0035 \\   Urban & 0.9735 & 0.9828 & 0.9925& 0.9851 & 0.9773 & 0.9971 & 0.0216 & 0.0037 \\   \hline \end{tabular} \end{center} \label{table1} \end{table*}

With the accuracy promise, we then investigate the robustness of the
learning models, i.e., the performance of the models with different
sizes of training data. We first split each sample set into two subsets,
one subset (occupying 10\% proportion) as validation set and the other
subset (occupying 90\% proportion) as training set. The training set
is evenly split into 10 subsets and for $j$-th training, for $j \in \{1,2,...10\}$,
the union of the first to $j$-th subsets behave as the training set.
Fig. \ref{fig:accu}(b) shows the accuracy of NLoS detection with
different sizes of training data under different scenarios. We have
the following two main observations. First, for NB method, to achieve
a very high accuracy, it requires high diversity training data to
cover all situations in validation set. When the accuracy performance
reaches a supreme value (about 96.5\% in the figure), increasing the
training size cannot further improve the performance. For instance,
in the highway scenario, the accuracy increases from 84.3\% to 90.9\%
then to 96.4\% with training data size 0.1, 0.2 and 0.3, respectively;
with more training data, the accuracy will oscillate around 96.4\%.
It is noted that different subsets of training data may have varied
impacts on the model performance, which explains that in the highway
scenario, the accuracy increases significantly with the 2nd and 3rd
subsets of training data, while in the suburban scenario, the accuracy
increase more obviously with the 5th subset of training data. Second,
SVM method is not as sensitive to the training data size as NB method
does. For instance, the accuracy of SVM in highway scenario oscillates
around 97\% regardless of the training data size. Similar observations
can also be obtained in suburban and urban scenarios. As the results
are tightly close to the highway results which may confuse the figure,
they are not shown in the figure.

\section{Conclusions\label{sec:Conclusions}}

In this article, we have discussed two important issues in VANETs
in the big data era, i.e., efficiently supporting the big data through
VANETs, and employing the big data to improve VANETs. For the former
one, a framework combining 5G cellular network and alternative opportunistic
data pipes is introduced, and is envisioned to provide efficient,
reliable, and flexible support of the VANETs big data. For the latter
one, the mechanisms which analyze and learn typical big data for characterizing
VANETs and designing intelligent protocols for VANETs are discussed.
Furthermore, we have shown a case study in which urban VANETs measurement
data is used to detect NLoS conditions through machine learning schemes.

\section{Acknowledgment}

This work is sponsored in part by the National Natural Science Foundation
of China (NSFC) under Grant No. 91638204, and the Natural Sciences
and Engineering Research Council of Canada.

\bibliographystyle{15_Users_YanpengDai_Desktop_Final_Files_IEEEtran}
\bibliography{14_Users_YanpengDai_Desktop_Final_Files_MyRefs}

\begin{thebibliography}{10}
\providecommand{\url}[1]{#1}
\csname url@samestyle\endcsname
\providecommand{\newblock}{\relax}
\providecommand{\bibinfo}[2]{#2}
\providecommand{\BIBentrySTDinterwordspacing}{\spaceskip=0pt\relax}
\providecommand{\BIBentryALTinterwordstretchfactor}{4}
\providecommand{\BIBentryALTinterwordspacing}{\spaceskip=\fontdimen2\font plus
\BIBentryALTinterwordstretchfactor\fontdimen3\font minus
  \fontdimen4\font\relax}
\providecommand{\BIBforeignlanguage}[2]{{%
\expandafter\ifx\csname l@#1\endcsname\relax
\typeout{** WARNING: IEEEtran.bst: No hyphenation pattern has been}%
\typeout{** loaded for the language `#1'. Using the pattern for}%
\typeout{** the default language instead.}%
\else
\language=\csname l@#1\endcsname
\fi
#2}}
\providecommand{\BIBdecl}{\relax}
\BIBdecl

\bibitem{al2014comprehensive}
S.~Al-Sultan, M.~M. Al-Doori, A.~H. Al-Bayatti, and H.~Zedan, ``A comprehensive
  survey on vehicular ad hoc network,'' \emph{Elsevier J. Netw. Comput. Appl.},
  vol.~37, pp. 380--392, 2014.

\bibitem{lu2014connected}
N.~Lu, N.~Cheng, N.~Zhang, X.~Shen, and J.~W. Mark, ``Connected vehicles:
  Solutions and challenges,'' \emph{{IEEE Internet Things J.}}, vol.~1, no.~4,
  pp. 289--299, 2014.

\bibitem{chen2014big}
M.~Chen, S.~Mao, and Y.~Liu, ``Big data: A survey,'' \emph{Springer Mobile
  Netw. Appl.}, vol.~19, no.~2, pp. 171--209, 2014.

\bibitem{bedi2014use}
P.~Bedi and V.~Jindal, ``Use of big data technology in vehicular ad-hoc
  networks,'' in \emph{Proc. IEEE ICACCI}, Delhi, India, 2014, pp. 1677--1683.

\bibitem{maurer2016autonomous}
M.~Maurer, J.~C. Gerdes, B.~Lenz, and H.~Winner, \emph{Autonomous driving:
  technical, legal and social aspects}.\hskip 1em plus 0.5em minus 0.4em\relax
  Springer, 2016.

\bibitem{hadded2015tdma}
M.~Hadded, P.~Muhlethaler, A.~Laouiti, R.~Zagrouba, and L.~A. Saidane,
  ``{TDMA-based MAC protocols for vehicular ad hoc networks: A survey,
  qualitative analysis, and open research issues},'' \emph{IEEE Commun. Surveys
  Tuts.}, vol.~17, no.~4, pp. 2461--2492, 2015.

\bibitem{quan2017enhancing}
W.~Quan, Y.~Liu, H.~Zhang, and S.~Yu, ``Enhancing crowd collaborations for
  software defined vehicular networks,'' \emph{IEEE Commun. Mag.}, vol.~55,
  no.~8, pp. 80--86, 2017.

\bibitem{andrews2014will}
J.~G. Andrews, S.~Buzzi, W.~Choi, S.~V. Hanly, A.~Lozano, A.~C. Soong, and
  J.~C. Zhang, ``What will 5g be?'' \emph{IEEE J. Sel. Areas Commun.}, vol.~32,
  no.~6, pp. 1065--1082, 2014.

\bibitem{3gppenhance}
3GPP, ``{3GPP TS 22.186: Enhancement of 3GPP support for V2X scenarios},''
  3GPP, Tech. Rep., June 2017.

\bibitem{bychkovsky2006measurementfull}
V.~Bychkovsky, B.~Hull, A.~Miu, H.~Balakrishnan, and S.~Madden, ``{A
  measurement study of vehicular Internet access using in situ Wi-Fi
  networks},'' in \emph{Proc. of ACM MobiCom}, Los Angeles, USA, Sep. 2006, pp.
  50--61.

\bibitem{cheng2013vehicular}
N.~Cheng, N.~Zhang, N.~Lu, X.~Shen, J.~Mark, and F.~Liu, ``{Opportunistic
  Spectrum Access for CR-VANETs: A Game-Theoretic Approach},'' \emph{IEEE
  Trans. Veh. Technol.}, vol.~63, no.~1, pp. 237--251, 2014.

\bibitem{zhou2017tv}
H.~Zhou, N.~Zhang, Y.~Bi, Q.~Yu, X.~Shen, D.~Shan, and F.~Bai, ``{TV} white
  space enabled connected vehicle networks: Challenges and solutions,''
  \emph{IEEE Netw.}, vol.~31, no.~3, pp. 6--13, 2017.

\bibitem{celes2017improving}
C.~Celes, F.~Silva, A.~Boukerche, R.~Andrade, and A.~Loureiro, ``Improving
  {VANET} simulation with calibrated vehicular mobility traces,'' \emph{IEEE
  Trans. Mobile Comput.}, vol.~16, no.~12, pp. 3376--3389, 2017.

\bibitem{mecklenbrauker2011vehicular}
C.~F. Mecklenbrauker, A.~F. Molisch, J.~Karedal, F.~Tufvesson, A.~Paier,
  L.~Bernad{\'o}, T.~Zemen, O.~Klemp, and N.~Czink, ``Vehicular channel
  characterization and its implications for wireless system design and
  performance,'' \emph{Proc. IEEE}, vol.~99, no.~7, pp. 1189--1212, 2011.

\bibitem{lv2016empirical}
F.~Lv, H.~Zhu, H.~Xue, Y.~Zhu, S.~Chang, M.~Dong, and M.~Li, ``An empirical
  study on urban ieee 802.11 p vehicle-to-vehicle communication,'' in
  \emph{Proc. IEEE SECON}, London, UK, Jun. 2016, pp. 1--9.

\end{thebibliography}

\section*{}

\begin{IEEEbiographynophoto}{Nan Cheng}
[S'12, M'16] received the Ph.D. degree from the Department of Electrical and Computer Engineering, University of Waterloo. He is currently working as a Post-doctoral fellow with the Department of Electrical and Computer Engineering, University of Toronto, and with Department of Electrical and Computer Engineering, University of Waterloo. His research interests include performance analysis and opportunistic communications for vehicular networks, unmanned aerial vehicles, and cellular traffic offloading.
\end{IEEEbiographynophoto}

\begin{IEEEbiographynophoto}{Feng Lyu}
 received the BS degree in software from Central South University
in 2013. He is pursuing his Ph.D degree in the Department of Computer
Science and Engineering at Shanghai Jiao Tong University. From October
2016 to October 2017, he was a visiting Ph.D. student at the Broadband
Communications Research (BBCR) group in the Department of Electrical
and Computer Engineering, University of Waterloo, Ontario, Canada.
His research interests include vehicular ad hoc networks and cloud
computing.
\end{IEEEbiographynophoto}

\begin{IEEEbiographynophoto}{Jiayin Chen}
 received the B.E. degree and the M.S. degree in the School of Electronics
and Information Engineering from Harbin Institute of Technology, Harbin,
China, in 2014 and 2016, respectively. She is currently pursuing the
Ph.D. degree with the Department of Electrical and Computer Engineering,
University of Waterloo, Waterloo, ON, Canada. Her research interests
are in the areas of vehicular communication networks, with current
focus on Intelligent Transport System, reinforcement learning, and
big data.
\end{IEEEbiographynophoto}

\begin{IEEEbiographynophoto}{Wenchao Xu}
received the B.E. and M.E. degrees from Zhejiang University, Hangzhou, China, in 2008 and 2011, respectively. He is currently working toward the Ph.D. degree with the Department of Electrical and Computer Engineering, University of Waterloo, Waterloo, ON, Canada. In 2011, he joined Alcatel Lucent Shanghai Bell Co. Ltd., where he was a Software Engineer for telecom virtualization. His interests include wireless communications with emphasis on resource allocation, network modeling, and mobile data offloading.
\end{IEEEbiographynophoto}

\begin{IEEEbiographynophoto}{Haibo Zhou}
[M'14, SM'18] received the Ph.D. degree in information and communication engineering
from Shanghai Jiao Tong University, Shanghai, China, in 2014. From
2014 to 2017, he has worked a Post-Doctoral Fellow with the Broadband
Communications Research Group, ECE Department, University of Waterloo.
Currently, he is an Associate Professor with the School of Electronic
Science and Engineering, Nanjing University. His research interests
include resource management and protocol design in cognitive radio
networks and vehicular networks.
\end{IEEEbiographynophoto}

\begin{IEEEbiographynophoto}{Shan Zhang}
[S'13, M'16] received her Ph.D. degree in Department of Electronic
Engineering from Tsinghua University, Beijing, China, in 2016. She
is currently an assistant professor in the Department of Computer
Science and Technology, Beihang University, Beijing, China. She was
a post doctoral fellow in Department of Electronical and Computer
Engineering, University of Waterloo, Ontario, Canada, from 2016 to
2017. Her research interests include mobile edge computing, wireless
network virtualization and intelligent management.
\end{IEEEbiographynophoto}

\begin{IEEEbiographynophoto}{Xuemin (Sherman) Shen}
[F] is a university professor, Department of Electrical and Computer Engineering, University of Waterloo, Canada. His research focuses on resource management, wireless network security, social networks, smart grid, and vehicular ad hoc networks. He is an IEEE Fellow, an Engineering Institute of Canada Fellow, a Canadian Academy of Engineering Fellow, and a Royal Society of Canada Fellow. He was a Distinguished Lecturer of the IEEE Vehicular Technology Society and the IEEE Communications Society.
\end{IEEEbiographynophoto}

\end{document}